# Thermal Conductivity and Spin State of the Spin Diamond-Chain System Azurite $Cu_3(CO_3)_2(OH)_2$

Yuta Hagiya[1], Takayuki Kawamata[1], Koki Naruse[1], Masumi Ohno[1], Yoshiharu Matsuoka[1], Hiroki Sudo[1], Hideki Nagasawa[1], Hikomitu Kikuchi[2], Takahiko Sasaki[3], and Yoji Koike[1]

[1] *Department of Applied Physics, Tohoku University, Sendai 980-8579, Japan*
[2] *Department of Applied Physics, Fukui University, Fukui 910-8507, Japan*
[3] *Institute for Materials Research (IMR), Tohoku University, Sendai 980-8577, Japan*

In order to investigate the spin state of azurite, $Cu_3(CO_3)_2(OH)_2$, we have measured the thermal conductivity along the *c*-axis, $\kappa_c$, perpendicular to the spin diamond-chains. It has been found that the temperature dependence of $\kappa_c$ shows a broad peak at ~ 100 K, which is explained as being due to the strong phonon-scattering by the strong spin-fluctuation owing to the spin frustration at low temperatures below ~ 100 K. Furthermore, it has been found that the temperature dependence of $\kappa_c$ shows another peak at low temperatures below 20 K and that $\kappa_c$ is suppressed by the application of magnetic field along the *c*-axis at low temperatures below ~ 35 K. In high magnetic fields above ~ 8 T at low temperatures below ~ 6 K, it has been found that $\kappa_c$ increases with increasing field. The present results have indicated that both spin-singlet dimers with a spin gap of ~ 35 K and antiferromagnetically correlated spin-chains with the antiferromagnetic exchange interaction of ~ 5.4 K are formed at low temperatures, which is consistent with the recent conclusion by Jeschke et al. [Phys. Rev. Lett. **106**, 217201 (2011)] that the ground state of spins in azurite in zero field is a spin-fluid one. In addition, a new quantum critical line in magnetic fields at temperatures above 3 K has been proposed to exist.

## 1. Introduction

The natural mineral azurite, $Cu_3(CO_3)_2(OH)_2$, attracts interest, because this compound is recognized as a model spin-system composed of geometrically frustrating spin diamond-chains.[1] As shown in Fig. 1, the crystal structure of azurite is monoclinic and consists of chains along the *b*-axis of alternately arranged monomers of Cu (called Cu1) and dimers of Cu (called Cu2). Since both Cu1 and Cu2 are $Cu^{2+}$ ions with the spin quantum number $S = 1/2$, distorted diamond-chains of spins with geometric frustration are formed along the *b*-axis. From the theoretical study,[2,3] three kinds of ground state in zero field have been proposed in spin distorted diamond-chain systems with geometric frustration, depending on the magnitude of the exchange interaction, $J_1$, $J_2$ and $J_3$, shown in Fig. 1. The magnitude of $J_1$, $J_2$ and $J_3$ of azurite, however, has been controversial both theoretically and experimentally.[4-10] Recently, the first principles density functional theory calculations by Jeschke et al.[11] have revealed that the characteristic experimental results in azurite[1], namely, both the so-called 1/3 plateau in the magnetization curve and the double-peak behavior in the temperature dependence of the magnetic susceptibility can be explained, taking into account the direct exchange interaction between Cu1 spins, $J_m$, shown in Fig. 1. The ground state of spins in azurite has proposed to be a spin-fluid one, where two spins of each Cu2 dimer are coupled by $J_2$ to form a singlet state and spins of Cu1 monomers are coupled by $J_m$ to form antiferromagnetically correlated spin-chains.[11]

It is known that thermal conductivity is a suitable probe to detect a change of the spin state in a spin system. In spin-gap systems, such as $CuGeO_3$,[12,13] $SrCu_2(BO_3)_2$[14-16] and $TlCuCl_3$,[17] an enhancement of the thermal conductivity due to phonon, $\kappa_{phonon}$, has been observed with decreasing temperature and the enhancement has been suppressed by the application of magnetic field at low temperatures below the temperature comparable to the spin-gap energy. In several antiferromagnets, a sudden enhancement of thermal conductivity has been observed just below the antiferromagnetic transition temperature.[18-23] These kinds of change of thermal conductivity occur due to the change of the scattering rate of heat carries, such as conduction electrons, phonons and magnetic excitations, sensitively affected by the change of the spin state. In azurite, in particular, changes of the spin state are expected to be detected from the change of $\kappa_{phonon}$, because the recent ultrasonic measurements have revealed that the magnetoelastic coupling is very strong.[24] Accordingly, in order to investigate the spin state of azurite, we have measured the thermal conductivity along the *c*-axis, $\kappa_c$, perpendicular to the spin diamond-chains, because $\kappa_{phonon}$ is dominant in $\kappa_c$ owing to little contribution of the thermal conductivity due to magnetic excitations, $\kappa_{spin}$, to $\kappa_c$.

## 2. Experimental

The single crystal of azurite was purchased at a stone shop.[1]  The quality of the single crystal is the same as that in the former paper,[1] for the temperature dependence of the magnetic susceptibility of the single crystal is almost the same as that in the former paper. Thermal conductivity measurements were carried out by the conventional steady-state method. One side of a rectangular single-crystal was anchored on the copper heat sink with indium solder.  A chip-resistance of 1 kΩ was attached as a heater to the opposite side of the single crystal with GE7031 varnish.  The temperature difference across the crystal was measured with two Cernox thermometers (Lake Shore Cryotronics CX-1050-SD).  The error of the absolute value of the thermal conductivity obtained was estimated to be about 10% on account of errors of the crystal geometry and the length between the temperature terminals. Magnetic fields up to 14 T were applied using a superconducting magnet.

## 3. Results and discussion

Figure 2 shows the temperature dependence of $\kappa_c$ in magnetic fields along to the $c$-axis up to 14 T.  In zero field, it is found that $\kappa_c$ shows a broad peak at ~ 100 K and a sharp peak at ~ 10 K.  Since azurite is a magnetic insulator, the thermal conductivity is given by the sum of $\kappa_{phonon}$ and $\kappa_{spin}$.  As mentioned in Sect. 1, however, little contribution of $\kappa_{spin}$ to $\kappa_c$ is expected, because the magnetic interaction along the $c$-axis perpendicular to the spin diamond-chains is very weak.  Therefore, the dominant contribution to $\kappa_c$ is $\kappa_{phonon}$.  It is noted that the typical temperature-dependence of $\kappa_{phonon}$ in a nonmagnetic insulator shows a peak around $\Theta_D/20$ ($\Theta_D$ : the Debye temperature), owing to the change of the dominant phonon-scattering from the phonon-phonon Umklapp scattering at high temperatures to the phonon-defect scattering at low temperatures.[25]  Since $\Theta_D$ has been estimated from the specific heat measurements as 188 K[8] and 350 K,[26] the appearance of a broad peak at ~ 100 K suggests that the mean free path of phonons, $l_{phonon}$, is limited by some additional phonon-scattering, which is inferred to be the strong phonon-scattering by the spin fluctuation owing to the frustration of $Cu^{2+}$ spins forming spin diamond-chains.  In several spin-frustrating systems, in fact, the thermal conductivity shows a broad peak at a high temperature far above $\Theta_D/20$ on account of the strong phonon-scattering by the spin

fluctuation.[27-29]

The sharp peak at ~ 10 K is explained as being due to the increase of $l_\text{phonon}$ on account of the decrease of the phonon-spin scattering rate owing to the decrease of the number of magnetic excitations with decreasing temperature at low temperatures below ~ 20 K, as in the case of spin-gap systems mentioned above.[12-17] That is, the sharp peak at ~ 10 K is due to the spin gap caused by the formation of a spin-singlet state of Cu2 dimers. Additionally, it is found that the sharp peak of $\kappa_c$ at ~ 10 K is suppressed by the application of magnetic field, which is reasonably explained as being due to the increase of the phonon-spin scattering rate owing to the reduction of the spin gap. The decrease of $\kappa_c$ by the application of magnetic field is observed at low temperatures below ~ 35 K, as shown in the insets (b) and (c) of Fig. 2. Therefore, both the spin-gap energy and $J_2$ are roughly estimated as ~ 35 K from the present thermal conductivity measurements. That is reasonable, because the spin-gap energy and $J_2$ in azurite has been estimated from the inelastic neutron scattering[8] and electron spin resonance[30] measurements as ~ 50 K.

At low temperatures below ~ 10 K, furthermore, it is found that $\kappa_c$ is markedly enhanced by the application of magnetic field along the $c$-axis, as shown in the insets (b) and (c) of Fig. 2. Figure 3 shows the magnetic-field dependence of the thermal conductivity along the $c$-axis normalized by the value in zero field, $\kappa_c(H)/\kappa_c(0)$, to see the change of $\kappa_c$ by the application of magnetic field clearly. It is found that $\kappa_c$ monotonically decreases with increasing field along the $c$-axis at temperatures between ~ 7 K and 18 K, owing to the reduction of spin gap with increasing filed. At low temperatures below ~ 6 K, on the other hand, $\kappa_c(H)/\kappa_c(0)$ markedly increases with increasing field in high fields above the characteristic field, $H_c$, ~ 8 T. That is, there is an enhanced contribution of thermal conductivity to $\kappa_c$ in high fields above $H_c$ ~ 8 T along the $c$-axis.[31] This is interpreted as follows. According to the conclusion by Jeschke et al.[11] in zero field at low temperatures, spins of Cu2 dimers form a spin-singlet state owing to $J_2$ ~ 33 K, while spins of Cu1 monomers form antiferromagnetically correlated spin-chains owing to $J_m$ ~ 4.6 K. In high magnetic fields above the field comparable to $k_B J_m/(g\mu_B)$ ($k_B$ : Boltzmann constant, $g$ : the g-factor, $\mu_B$ : the Bohr magneton) at low temperatures, on the other hand, it has been pointed out from the NMR measurements that the spin-singlet state of Cu2 dimers is maintained, while spins of Cu1 monomers are completely polarized in the field direction, leading to the appearance of the 1/3 plateau in the magnetization curve.[26,32] Since the field-induced ferromagnetic state of Cu1 spins is expected to hardly scatter phonons, $l_\text{phonon}$ increases and $\kappa_c(H)/\kappa_c(0)$ is enhanced in high magnetic fields above $H_c \sim 8$ T.   The field above which the ferromagnetic state is formed, namely, the field above which the 1/3 plateau is observed in the magnetization curve has been estimated from the ultrasonic measurements at low temperatures below 2 K, as shown in Fig. 4.[33])   It is found that this field, located at the quantum critical line between the 1/3 plateau state and the paramagnetic state, is ~ 10 T at 2 K and increases with decreasing temperature.   When values of $H_c$, defined at the intersection point of two extrapolated lines of the fitted line of $\kappa_c(H)/\kappa_c(0)$ vs $H$ in low fields and the steepest part of $\kappa_c(H)/\kappa_c(0)$ vs $H$ in high fields, at various temperatures are plotted in Fig. 4, it is found that they are smoothly connected to the quantum critical line.   Accordingly, this interpretation of the magnetic-field dependence of $\kappa_c$ is confirmed to be reasonable.   At high temperatures above 7 K, it is hard to determine $H_c$ on account of the little increase in $\kappa_c(H)/\kappa_c(0)$, indicating that the field-induced ferromagnetic state of Cu1 spins is disturbed due to the thermal fluctuation.   In fact, the 1/3 plateau in the magnetization curve is smeared at 4.2 K.[1)]   From the present results, moreover, $J_m$ is roughly estimated as ~ 5.4 K using the equation, $J_m = g\mu_B SH_c/k_B$.   This value is comparable to $J_m \sim 4.6$ K by Jeschke et al.[11)]   The present results support the conclusion by Jeschke et al.[11)] that the ground state of spins in azurite in zero field is a spin-fluid one.   In addition, a new quantum critical line has been obtained at temperatures above 3 K.   It attracts interest that this quantum critical line is confirmed by other measurements.

## 4.  Summary

     We have measured $\kappa_c$ perpendicular to the spin diamond-chains of azurite to investigate the spin state.   It has been found that the temperature dependence of $\kappa_c$ shows a broad peak at ~ 100 K, which is due to the strong phonon-scattering by the strong spin-fluctuation owing to the spin frustration at low temperatures below ~ 100 K.   Furthermore, it has been found that the temperature dependence of $\kappa_c$ shows another peak at low temperatures below 20 K and that $\kappa_c$ is suppressed by the application of magnetic field at low temperatures below ~ 35 K.   These indicate the existence of a spin gap of ~ 35 K, because the enhancement of $\kappa_c$ is due to the increase of $\kappa_{phonon}$ which is caused by the suppression of the phonon-spin scattering owing to the spin-gap formation and because the suppression with increasing field is due to the reduction of the spin gap.   In high magnetic fields above ~ 8 T at low temperatures below ~ 6 K, it has been found that $\kappa_c$ increases with increasing fields.   This has been explained as being due to the decrease of the phonon-spin scattering rate owing to the formation of a

ferromagnetic state of Cu1 spins induced by the application of high magnetic fields above the field comparable to $k_B J_m/(g\mu_B)$. The value of $J_m$ has roughly been estimated as ~ 5.4 K. The present results of $\kappa_c$ are consistent with the conclusion by Jeschke et al.[11] that the ground state of spins in azurite in zero field is a spin-fluid one with singlet dimers of Cu2 spins coupled by $J_2$ and antiferromagnetic spin-chains of Cu1 spins coupled by $J_m$. In addition, a new quantum critical line between the paramagnetic state and the 1/3 plateau state in magnetic fields at temperatures above 3 K has been proposed to exist.

**Acknowledgments**

Fig.1

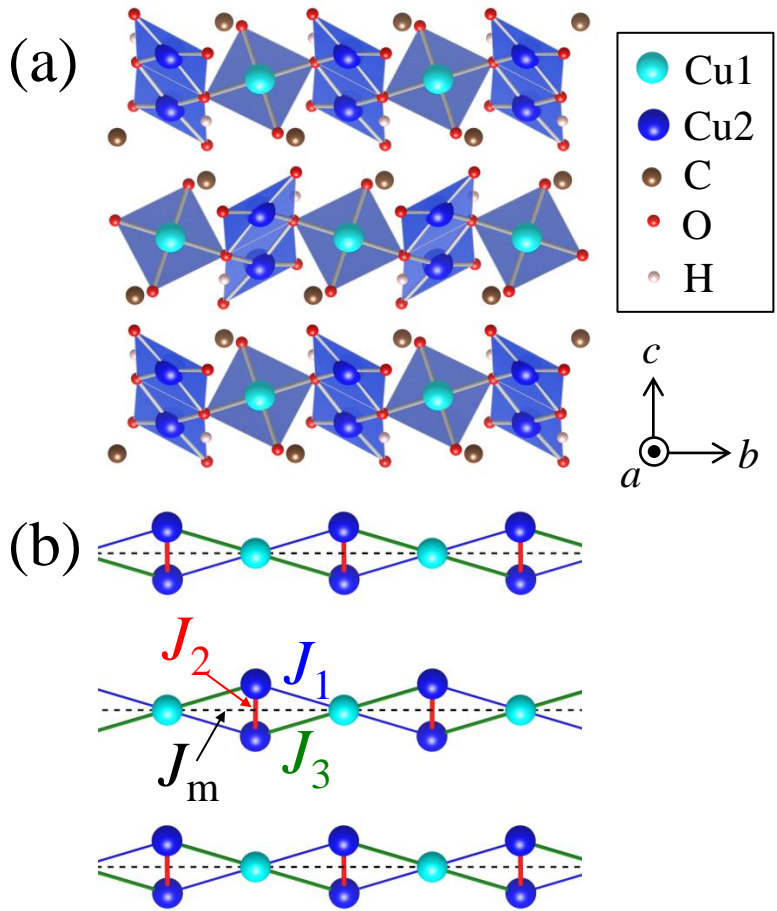

Schematic views of (a) the crystal structure and (b) spin diamond-chains of azurite, $Cu_3(CO_3)_2(OH)_2$. Spins of Cu1 and Cu2 with spin-1/2 interact with the exchange interaction $J_1$, $J_2$, $J_3$ and $J_m$. It is noted that the $c$-axis is not in the paper plane but slightly tilted to the direction of the depth of the paper plane because of its monoclinic structure.

Fig.2

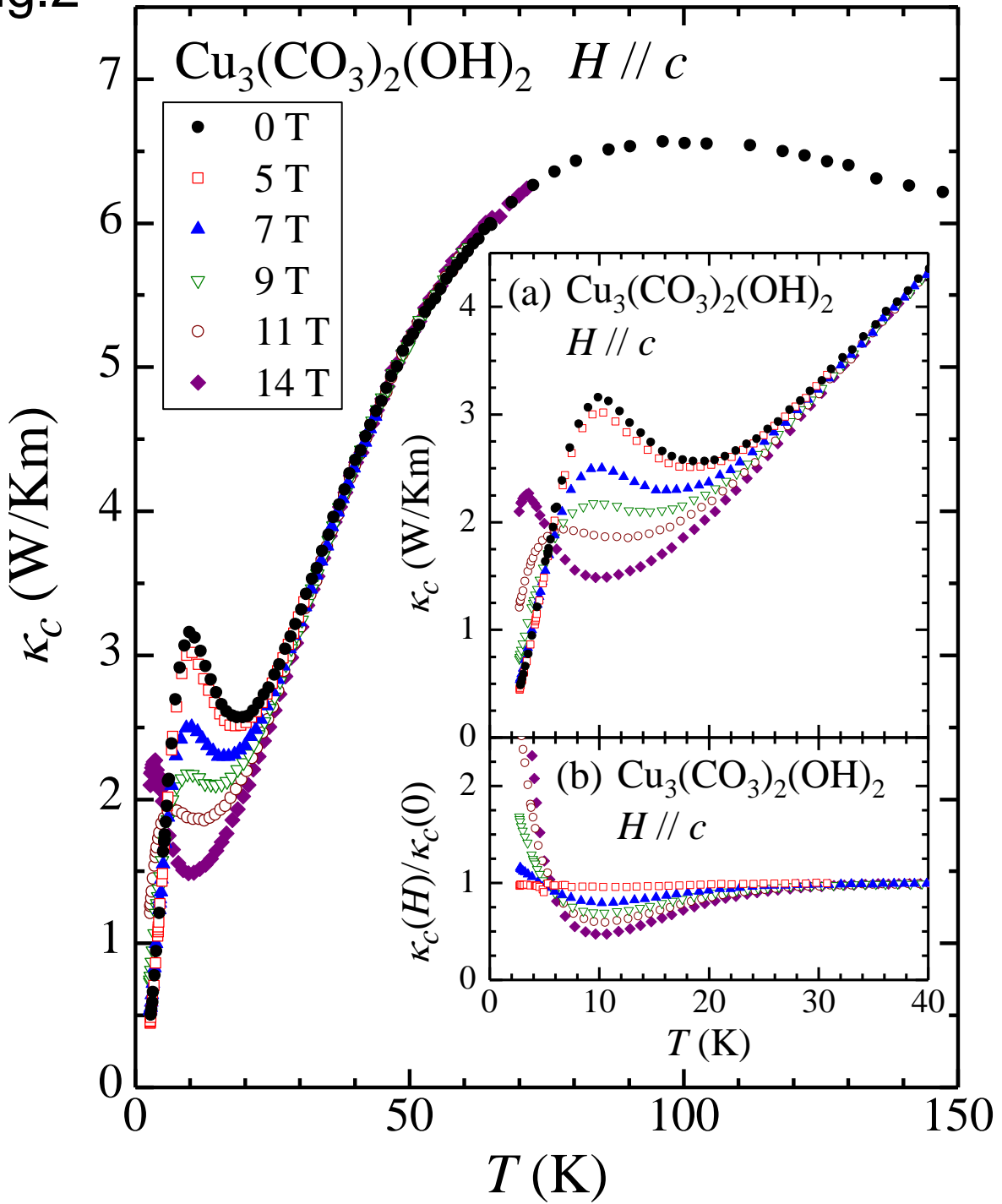

Temperature dependence of the thermal conductivity along the $c$-axis, $\kappa_c$, perpendicular to the spin diamond-chains of azurite, $Cu_3(CO_3)_2(OH)_2$, in magnetic fields along the $c$-axis. The inset (a) is an expanded plot. The inset (b) shows the temperature dependence of $\kappa_c$ normalized by the value in zero filed, $\kappa_c(H)/\kappa_c(0)$, in magnetic fields along the $c$-axis.

Fig.3

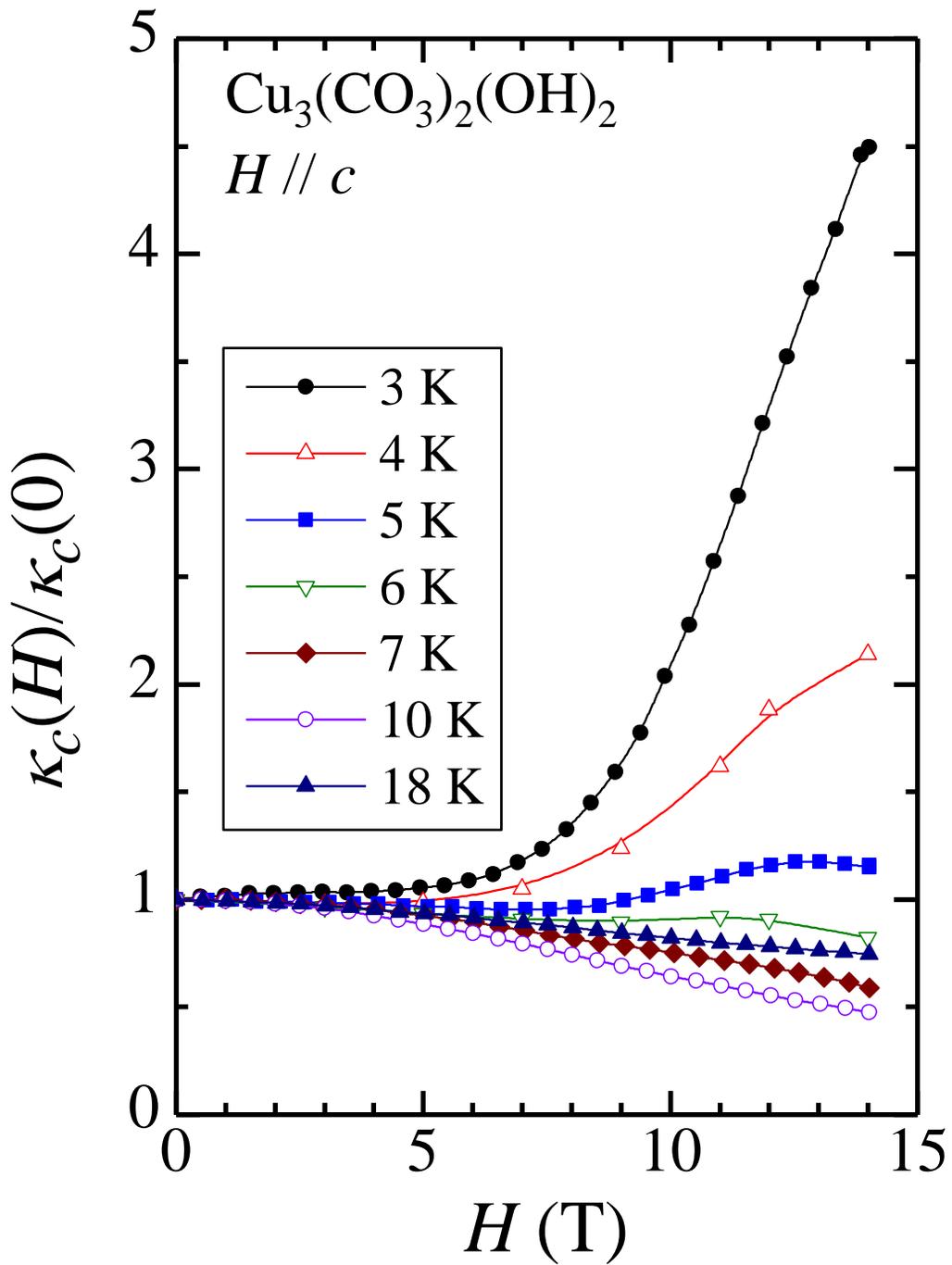

Magnetic-field dependence of the thermal conductivity along the *c*-axis normalized by the value in zero filed, $\kappa_c(H)/\kappa_c(0)$, perpendicular to the spin diamond-chains of azurite, $Cu_3(CO_3)_2(OH)_2$, in magnetic fields along the *c*-axis.

Fig.4

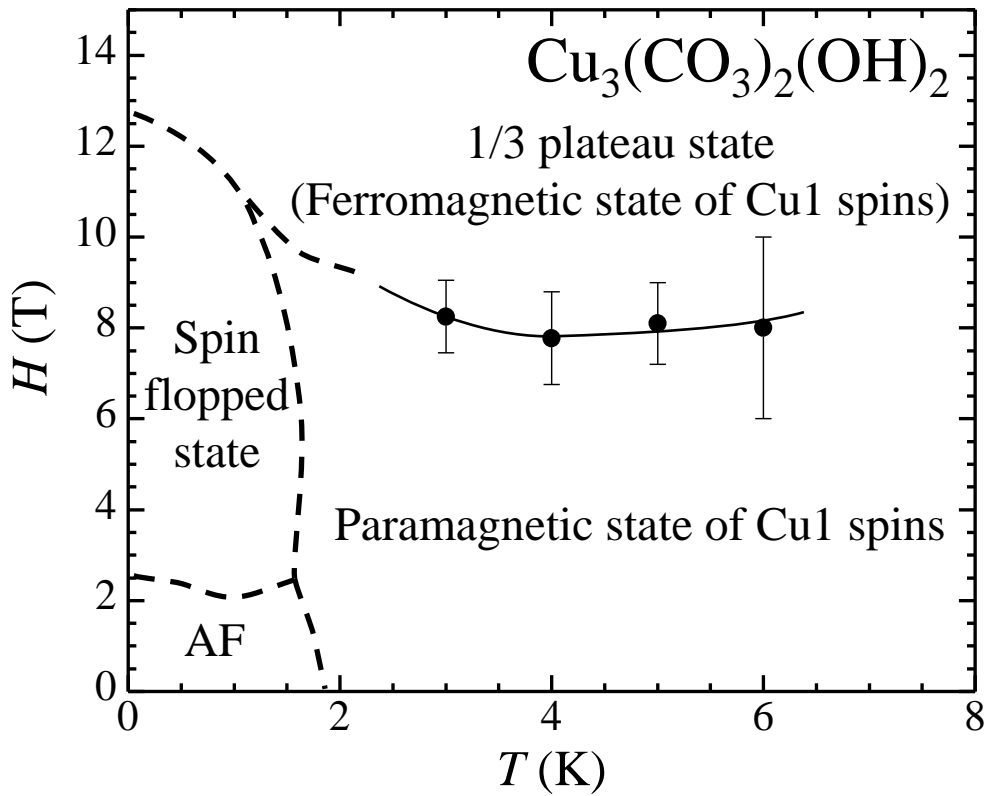

Magnetic field $H$ vs. temperature $T$ phase diagram of azurite, $Cu_3(CO_3)_2(OH)_2$. Solid circles are estimated from the magnetic-field dependence of the thermal conductivity along the $c$-axis, $\kappa_c$, in magnetic fields parallel to the $c$-axis. The solid line indicates a guide for the eyes. The broken line indicates boundaries estimated from the ultrasonic measurements in magnetic fields perpendicular to the $b$-axis.[33] AF indicates the antiferromagnetic state.